\documentstyle[aps,preprint]{revtex}
\tighten
\begin{document}
\draft


\preprint{gr-qc/0009014}
\title{Moving-mirror entropy}
\author{Shinji Mukohyama and Werner Israel}
\address{
Department of Physics and Astronomy, University of Victoria\\ 
Victoria, BC, Canada V8W 3P6
}
\date{\today}

\maketitle


\begin{abstract} 

We consider a quantized scalar field in a two-dimensional Minkowski
spacetime with a moving mirror and propose a definition of
moving-mirror entropy associated with temporarily inaccessible
information about the future. 

\end{abstract}




One of the most embarrassing problems in gravitational physics is the
so called information loss problem. When Hawking discovered a thermal
radiation from black holes (Hawking radiation) in 1974 by using
semiclassical and quasi-stationary approximations, he argued that 
formation and evaporation of black holes would introduce a
non-unitarity in quantum mechanics~\cite{Hawking}. Actually, his 
calculation shows that an initial pure state evolves to a totally
uncorrelated thermal state, which is impossible in a unitarily
evolving system. In general within these approximations, the evolution 
is described by a non-unitary superscattering matrix which maps
initial mixed states to final mixed states, and superscattering matrix
elements can be calculated
explicitly~\cite{Panangaden&Wald,Mukohyama1997}. Hence, as far as the 
approximations are justified, some information with respect to the
initial state seems to be lost.


On the other hand, Page~\cite{Page1980} argued that, if the
superscattering matrix describing the whole system including quantum
gravity is CPT invariant, the description can be reduced to an $S$
matrix which maps pure initial states into pure final states. In other
words, if quantum gravity is CPT invariant then there is no loss of
information in the process of formation and evaporation of black
holes. In this case information can be lost only temporarily and the
temporarily missing information should be completely recovered. 
Although his arguments might seem to contradict the semiclassical 
result that emission from a black hole is uncorrelated, he also argued 
later that information may come out initially so slowly, or else be so
spread out, that it would never show up in a perturbative
analysis~\cite{Page1993}. Hence, it seems very difficult to prove or
disprove his arguments by any perturbative analysis, in particular the
semiclassical calculation.


Page's arguments seem to be consistent with the interpretation of
entanglement entropy proposed in \cite{Mukohyama1998,Mukohyama2000}. 
Entanglement entropy of a pure state with respect to a division of
a Hilbert space into two subspaces $1$ and $2$ was interpreted as an
amount of information which can be transmitted through $1$ and $2$
from a system interacting with $1$ to another system interacting with
$2$. In this interpretation, the transmitting medium is the quantum
entanglement between $1$ and $2$, and the entanglement entropy may be 
a quantity which can in principle cancel the black-hole entropy, the 
amount of the temporarily missing information, to restore information
loss. However, so far we do not have any concrete models to realize
this idea.


Recent progress in string theory suggests the existence of a unitary
$S$-matrix description for the Hawking process. (For reviews of black
holes in superstring theory see, for example,
\cite{Horowitz,Maldacena,Peet} and references therein.) In the D-brane
picture, an extremal or near extremal black hole is described by
stacked D-branes on which open strings are attached and
moving. Hawking radiation can be described as a decay process of two
open strings into a closed string away from the D-branes. Since the
evolution of the whole system in the D-brane picture including closed
and open strings is manifestly unitary, the Hawking process should be 
described as a unitary process at least for near extremal black
holes. Moreover, the AdS/CFT
correspondence~\cite{Maldacena-AdS/CFT,AdS/CFT} might be considered
as a conjecture about a unitary description for arbitrary
gravitational processes including formation and evaporation of black
holes.


The D-brane picture also provides us an interpretation of black-hole
entropy of an extremal or near extremal black hole as the logarithm
of the number of different states of open strings on stacked
D-branes~\cite{Strominger&Vafa}. Hence, in some sense the black-hole entropy seems to be an 
amount of temporarily inaccessible information. Namely, if the black
hole evaporates completely then information about the quantum state of
the black hole should come out as correlations among closed strings,
provided that a unitary description similar to the D-brane picture
holds for evaporating black holes.


However, it seems fair to say that the above consideration about
the restoration of missing information is no more than
speculation. For example, see Hawking's
objections~\cite{Hawking1998}. Moreover, because of the Page's 
arguments that the recovery of information would never show up in a 
perturbative analysis, it seems very difficult to prove or disprove
it.

On the other hand, one might consider a different approach: it might
be relevant to investigate whether it is possible to assign entropy to
a manifestly unitary process. In this paper, we shall consider the so 
called moving-mirror effect~\cite{Fulling,Davies}, which is manifestly 
unitary but is often referred as an analogue of the Hawking process.


For simplicity, consider a minimally-coupled scalar field in
two-dimensional Minkowski spacetime ($ds^2=-dt^2+dx^2=-dudv$) with a 
moving-mirror. Denote the trajectory of the mirror by
$x^{\mu}=X^{\mu}(\tau)$, where $\tau$ is the proper time along the 
mirror trajectory and assume that both sides of the mirror are
perfectly reflecting. For this situation, suppose that the
functions $X^{\mu}(\tau)$ for $\tau<\tau_0$ are known and that the
remaining $\tau>\tau_0$ part of the trajectory is unknown. Of course,
provided that the initial state at ${\cal I}^-$ is given, we can
determine the quantum state of wave-packet modes which are reflected
by the mirror before $\tau_0$. However, we cannot predict the quantum
state of wave-packet modes which will be reflected by the mirror after
$\tau_0$ since that depends on the unknown future trajectory of the
mirror. In the following, we shall propose a definition of
moving-mirror entropy associated with the lack of information about
the future trajectory of the mirror.


It is expected that this lack of information should be related to
uncertainty about the quantum state of wave-packet modes which will be 
reflected by the mirror in the future since the latter is determined
by the former, if the initial quantum state at ${\cal I}^-$ is
fixed. In order to represent such uncertainty quantitatively, it is
convenient to consider a quantum-mechanical generalization of
conditional entropy, the so called von Neumann conditional entropy.

Classically, the conditional entropy of an experiment $A$ relative to
another experiment $B$ is defined by 
$H(A|B) = -\sum_{a,b}p(a,b)\ln p(a|b)$, where $a$ and $b$ represent
outcomes of $A$ and $B$, respectively, $p(a,b)$ is the joint
probability of $a$ and $b$, $p(a|b)=p(a,b)/p(b)$ is the conditional
probability of $a$ for a given outcome $b$, and $p(b)$ is the
probability of $b$. The conditional entropy corresponds to uncertainty
about the outcome of $A$ after the experiment $B$ has been
performed. The quantum analogue of conditional entropy was considered
in references~\cite{Cerf&Adami1,Cerf&Adami2}. Consider a Hilbert space
${\cal F}$ of the form ${\cal F}={\cal F}_1\otimes{\cal F}_2$ and let
$\rho$ be a density matrix on $\cal{F}$. Given $\rho$, the conditional
entropy of the subsystem $1$ relative to $2$ is defined by 
%
\begin{equation}
 S_{1|2} = {\bf Tr}\left[\rho\sigma_{1|2}\right],
\end{equation}
where $\sigma_{1|2}={\bf 1}_{1}\otimes\ln\rho_2-\ln\rho$, and 
%
\begin{equation}
 \rho_2={\bf Tr}_1\rho. 
\end{equation}

In our case, ${\cal F}_1$ and ${\cal F}_2$ are symmetric Fock
spaces constructed from Hilbert spaces ${\cal H}_1$ and 
${\cal H}_2$ of mode functions which are reflected by the mirror
after and before $\tau_0$, respectively. 
It seems natural to assume that the initial quantum
state is the vacuum $|0,-\rangle$ determined by positive-frequency mode
functions at ${\cal I}^-$. Correspondingly, the density matrix $\rho$
represents a pure state given by $\rho=|0,-\rangle\langle 0,-|$. It is
easy to see that, if $\rho$ is pure, 
%
\begin{equation}
 S_{1|2} = -S_{ent},
\end{equation}
where $S_{ent}$ is entanglement entropy defined by
%
\begin{eqnarray}
 S_{ent} = -{\bf Tr}_2[\rho_2\ln\rho_2].
\end{eqnarray}
Moreover, in this case, the quantum state $|0,-\rangle$ is the direct
product of quantum states $|0,-R\rangle$ and $|0,-L\rangle$ in the
right and left sectors R and L separated by the mirror. 
Hence, the entanglement entropy can be calculated separately in each
of the two regions, and the total entanglement entropy is the sum of
these two contributions.


To calculate the entanglement entropy in the R region, it is
convenient to write the mirror trajectory as $v=p(u)$ and to introduce
a new coordinate $\tilde{u}$ by $\tilde{u}\equiv p(u)$. In terms of
this coordinate, the mirror trajectory becomes $v=\tilde{u}$ and the R
region is $v\ge\tilde{u}$. Thence, as we shall briefly illustrate
below, the entanglement entropy in the R region is easily calculated
and given by 
%
\begin{equation}
 S_{entR} = \frac{1}{12}\ln [L_{\tilde{u}}/l_{\tilde{u}}],
\end{equation}
where $L_{\tilde{u}}$ and $l_{\tilde{u}}$ are infrared and
ultraviolet cutoffs in the $\tilde{u}$-coordinate~\cite{FPST}.

Now let us verify this expression for the case of Dirichlet boundary
conditions at the mirror. For this purpose note first that the
entanglement entropy does not depend on the bases in ${\cal F}_{1}$
and ${\cal F}_{2}$ and so we can choose these bases in a convenient
way. Hence, following Unruh~\cite{Unruh}, we introduce Rindler
coordinates $\tilde{u}^{(1)}$ and $\tilde{u}^{(2)}$ defined by 
%
\begin{eqnarray}
 \tilde{u}^{(1)} & = & 
	-\ln (\tilde{u}-\tilde{u}_0)\quad\mbox{for}
	\quad \tilde{u} > \tilde{u}_0, \nonumber\\
 \tilde{u}^{(2)} & = & 
	-\ln (-\tilde{u}+\tilde{u}_0)\quad\mbox{for}
	\quad \tilde{u} < \tilde{u}_0,
\end{eqnarray}
where $\tilde{u}_0=p(X^u(\tau_0))$. Next, we can construct symmetric
Fock spaces ${\cal F}_{1R}$ and ${\cal F}_{2R}$ from the following
mode functions $\varphi^{(1)}_{\omega}$ and $\varphi^{(2)}_{\omega}$,
respectively. 
%
\begin{eqnarray}
 \varphi^{(1)}_{\omega} & = &
	\theta(\tilde{u}-\tilde{u}_0)e^{i\omega\tilde{u}^{(1)}}
	- \theta(v-\tilde{u}_0)e^{i\omega v^{(1)}}, \nonumber\\
 \varphi^{(2)}_{\omega} & = & 
	\theta(-\tilde{u}+\tilde{u}_0)e^{-i\omega\tilde{u}^{(2)}}
	- \theta(-v+\tilde{u}_0)e^{-i\omega v^{(2)}},
\end{eqnarray}
where $v^{(1)}$ and $v^{(2)}$ are defined by 
%
\begin{eqnarray}
 v^{(1)} & = & 
	-\ln (v-\tilde{u}_0)\quad\mbox{for}
	\quad v > \tilde{u}_0, \nonumber\\
 v^{(2)} & = & 
	-\ln (-v+\tilde{u}_0)\quad\mbox{for}
	\quad v < \tilde{u}_0. 
\end{eqnarray}
Evidently, the Fock space ${\cal F}_{1R}$ (and ${\cal F}_{2R}$) is the 
subspace of ${\cal F}_1$ (and ${\cal F}_2$, respectively) which is
relevant to the R region. Thirdly, the vacuum $|0,-R\rangle$ at 
${\cal I}^-$ in the R region can be defined by the following set of
mode functions $\Phi^{(1)}_{\omega}$ and $\Phi^{(2)}_{\omega}$ which
are analytic in the lower complex $v$-plane. 
%
\begin{eqnarray}
 \Phi^{(1)}_{\omega} & = &
	N_{\omega}(\varphi^{(1)}_{\omega}
	+ e^{-\pi\omega}\varphi^{(2)*}_{\omega}), \nonumber\\
 \Phi^{(2)}_{\omega} & = & 
	N_{\omega}(\varphi^{(2)}_{\omega}
	+ e^{-\pi\omega}\varphi^{(1)*}_{\omega}),
\end{eqnarray}
where $N_{\omega}$ is a normalization constant. Thence, following
\cite{Israel}, it is now easy to expand the vacuum $|0,-R\rangle$ in
terms of Fock space states constructed from
$\{\varphi^{(1,2)}_{\omega}\}$ and to trace over the degrees of
freedom in ${\cal F}_{1R}$. The resulting reduced density matrix on 
${\cal F}_{2R}$ is the thermal density matrix with temperature
$T=1/2\pi$, provided that it is written in terms of Fock space states
constructed from $\{\varphi^{(2)}_{\omega}\}$. Therefore the 
corresponding entropy density, or entropy per unit interval of the
$\tilde{u}^{(2)}$ coordinate, is ${\cal S}=(\pi/6)T=1/12$.  Finally,
by introducing the ultraviolet and infrared cutoffs $l_{\tilde{u}}$
and $L_{\tilde{u}}$, we obtain  
%
\begin{equation}
 S_{entR} = \int_{\tilde{u}=\tilde{u}_0-L_{\tilde{u}}}
	^{\tilde{u}=\tilde{u}_0-l_{\tilde{u}}}
	{\cal S}d\tilde{u}^{(2)}
	= \frac{1}{12}\ln[L_{\tilde{u}}/l_{\tilde{u}}].
\end{equation}

Similarly, to calculate the entanglement entropy in the L region,
we write the mirror trajectory as $u=q(v)$ and introduce a new
coordinate $\tilde{v}$ by $\tilde{v}\equiv q(v)$. Using this
coordinate, the entanglement entropy in the L region is given by 
%
\begin{equation}
 S_{entL} = \frac{1}{12}\ln [L_{\tilde{v}}/l_{\tilde{v}}],
\end{equation}
where $L_{\tilde{v}}$ and $l_{\tilde{v}}$ are infrared and
ultraviolet cutoffs in the $\tilde{v}$-coordinate.

The ultraviolet cutoffs $l_{\tilde{u}}$ and $l_{\tilde{v}}$ can be
related to the ultraviolet cutoff $l$ in the proper time
$\tau$ as follows. 
%
\begin{eqnarray}
 l_{\tilde{u}} & = & \frac{d\tilde{u}}{d\tau}l
	+\frac{1}{2}\frac{d^2\tilde{u}}{d\tau^2}l^2 
	+\frac{1}{6}\frac{d^3\tilde{u}}{d\tau^3}l^3 
	+ O(l^4),
	\nonumber\\
 l_{\tilde{v}} & = & \frac{d\tilde{v}}{d\tau}l
	+\frac{1}{2}\frac{d^2\tilde{v}}{d\tau^2}l^2 
	+\frac{1}{6}\frac{d^3\tilde{v}}{d\tau^3}l^3 
	+ O(l^4),
\end{eqnarray}
or
%
\begin{eqnarray}
 l_{\tilde{u}} & = & (p'(u))^{1/2}l
	\left[ 1+\frac{1}{2}la^{\mu}n_{\mu} 
	+\frac{l^2}{6}\left(\frac{d}{d\tau}(a^{\mu}n_{\mu})
	+a^2\right)
	+ O(l^3a^3,l^3a\partial_{\tau}a)\right], 
	\nonumber\\
 l_{\tilde{v}} & = & (q'(v))^{1/2}l
	\left[ 1-\frac{1}{2}la^{\mu}n_{\mu} 
	+\frac{l^2}{6}\left(-\frac{d}{d\tau}(a^{\mu}n_{\mu})
	+a^2\right)
	+ O(l^3a^3,l^3a\partial_{\tau}a)\right],
\end{eqnarray}
where $a^{\mu}=d^2X^{\mu}/d\tau^2$, $a^2=a^{\mu}a_{\mu}$, and
$n^{\mu}$ is the unit normal to the trajectory directed toward the
R region. Since $p'q'=1$, we obtain 
%
\begin{equation}
 l_{\tilde{u}}l_{\tilde{v}} = 
	l^2\left[ 1+\frac{1}{12}l^2a^2 
	+ O(l^3a^3,l^3a\partial_{\tau}a)\right]
\end{equation}
as a function of the proper time $\tau$ along the mirror
trajectory. On the other hand, provided that the mirror is
asymptotically inertial, we obtain $L_{\tilde{u}}L_{\tilde{v}}=L^2$,
where $L$ is the infrared cutoff in the proper time $\tau$. Hence, we
obtain the total entanglement entropy. 
%
\begin{equation}
 S_{ent} = S_{entR}+S_{entL} = 
	-\frac{1}{144}l^2a^2 
	\left[1+O(la,l\partial_{\tau}\ln a)\right]
	+ \frac{1}{6}\ln[L/l]. 
\end{equation}
Thus, 
%
\begin{equation}
 S_{1|2} = \frac{1}{144}l^2a^2 
	\left[1+O(la,l\partial_{\tau}\ln a)\right]
	- \frac{1}{6}\ln[L/l]. 
\end{equation}


Now we define {\it moving-mirror entropy} so that it measures how
much the motion of the mirror increases the uncertainty of the quantum
state of wave-packet modes which are reflected by the mirror after
$\tau_0$. The simplest procedure is to subtract $S_{1|2}$ for a
non-accelerating mirror from $S_{1|2}$ for the actual trajectory:
%
\begin{equation}
 S_{MM}(\tau_0) \equiv S_{1|2} - S_{1|2}|_{a=0}
	= \frac{1}{144}l^2a^2(\tau_0)
	\left[ 1 + O(la,l\partial_{\tau}\ln a)\right].
	\label{eqn:Smoving}
\end{equation}


In summary we have calculated von Neumann conditional entropy for a
quantized scalar field in a two-dimensional Minkowski spacetime with a
moving mirror and proposed a definition of moving-mirror entropy
$S_{MM}$ as the difference between von Neumann conditional entropy for
the actual mirror trajectory and the inertial trajectory. We expect
that $S_{MM}$ is related to uncertainty about the future or
temporarily inaccessible information. Since black-hole entropy has
many interpretations~\cite{Frolov&Fursaev,Iyer&Wald,Mukohyama-thesis}, 
probably it is also worth while investigating other possible
interpretations of the moving-mirror entropy.

\vspace{1cm}

One of us (S.M.) is supported by CITA National Fellowship and the 
NSERC operating research grant. The other (W.I.) is supported by the
Canadian Institute for Advanced Research and the NSERC operating
research grant.


\end{document}